\documentclass[aps,prl,floatfix,twocolumn,byrevtex,superscriptaddress]{revtex4-1}
\usepackage[english]{babel}
\usepackage{amsmath}
\usepackage{graphicx}
\usepackage{float}
\usepackage{bm}
\usepackage{multirow}
\usepackage{dcolumn}
\usepackage{hyperref}

\newcommand{\icm}{\ensuremath{~\textrm{cm}^{-1}}}
\newcommand{\BKFAx}{Ba$_{1-x}$K$_{x}$Fe$_{2}$As$_{2}$}

\begin{document}

\title{Scaling of the Fano effect of the in-plane Fe-As phonon and the superconducting critical temperature in Ba$_{1-x}$K$_{x}$Fe$_{2}$As$_{2}$}

\author{B. Xu}
\email[]{bing.xu@unifr.ch}
\affiliation{University of Fribourg, Department of Physics and Fribourg Center for Nanomaterials, Chemin du Mus\'{e}e 3, CH-1700 Fribourg, Switzerland}

\author{E. Cappelluti}
\affiliation{
Istituto di Struttura della Materia, CNR, 34149 Trieste, Italy}

\author{L. Benfatto}
\affiliation{ISC-CNR and Department of Physics, Sapienza University of Rome, P.le A. Moro 5, 00185 Rome, Italy}

\author{B. P. P. Mallett}
\affiliation{University of Fribourg, Department of Physics and Fribourg Center for Nanomaterials, Chemin du Mus\'{e}e 3, CH-1700 Fribourg, Switzerland}
\affiliation{The Photon Factory, Department of Physics, University of Auckland, 38 Princes St, Auckland, New Zealand}

\author{P. Marsik}
\author{E. Sheveleva}
\author{F. Lyzwa}
\affiliation{University of Fribourg, Department of Physics and Fribourg Center for Nanomaterials, Chemin du Mus\'{e}e 3, CH-1700 Fribourg, Switzerland}

\author{Th. Wolf}
\affiliation{Institute of Solid State Physics, Karlsruhe Institute of Technology, Postfach 3640, Karlsruhe 76021, Germany}

\author{R. Yang}
\author{X. G. Qiu}
\affiliation{Beijing National Laboratory for Condensed Matter Physics, Institute of Physics, Chinese Academy of Sciences, Beijing 100190, China}

\author{Y. M. Dai}
\author{H. H. Wen}
\affiliation{National Laboratory of Solid State Microstructures and Department of Physics, Nanjing University, Nanjing 210093, China}

\author{R. P. S. M. Lobo}
\affiliation{LPEM, ESPCI Paris, PSL University, CNRS, F-75005 Paris, France}
\affiliation{Sorbonne Universit\'e, CNRS, LPEM, F-75005 Paris, France}

\author{C. Bernhard}
\email[]{christian.bernhard@unifr.ch}
\affiliation{University of Fribourg, Department of Physics and Fribourg Center for Nanomaterials, Chemin du Mus\'{e}e 3, CH-1700 Fribourg, Switzerland}

\date{\today}

%
%
\begin{abstract}
By means of infrared spectroscopy we determine the temperature-doping phase diagram of the Fano effect for the in-plane Fe-As stretching mode in Ba$_{1-x}$K$_{x}$Fe$_{2}$As$_{2}$. The Fano parameter $1/q^2$, which is a measure of the phonon coupling to the electronic particle-hole continuum, shows a remarkable sensitivity to the magnetic/structural orderings at low temperatures. More strikingly, at elevated temperatures in the paramagnetic/tetragonal state we find a linear correlation between $1/q^2$ and the superconducting critical temperature $T_c$. Based on theoretical calculations and symmetry considerations, we identify the relevant interband transitions that are coupled to the Fe-As mode. In particular, we show that a sizable $xy$ orbital component at the Fermi level is fundamental for the Fano effect and possibly also for the superconducting pairing.
\end{abstract}



\maketitle

%
%
The identification of the superconducting (SC) pairing mechanism of the iron arsenides is complicated by their multi-band and multi-gap structure and by the entanglement of the magnetic, orbital and structural degrees of freedom~\cite{Paglione2010,Stewart2011,Mazin2010,Wang2011,Chubukov2012,Medici2014,Fernandes2014}. A prominent example is the stripe-like antiferromagnetic (AF) order in undoped and weakly doped \BKFAx\ (BKFA), described by a single $q$-vector along (0, $\pi$), that is accompanied by an orthorhombic distortion of the high temperature tetragonal structure with strong nematic fluctuations~\cite{Fernandes2014}. This orthorhombic AF (o-AF) order persists well into the SC regime where it competes with SC at $0.15 \leq x \leq 0.3$. The strongest SC response, in terms of the $T_c$ value~\cite{Avci2014,Bohmer2015}, condensate density~\cite{Mallett2017PRB} and condensation energy~\cite{Storey2013}, occurs around optimum doping ($x$ = 0.3 -- 0.35) just as the o-AF order vanishes. Shortly before this point, around $x$ = 0.24 -- 0.26, a different AF order occurs that has the spins reoriented along the c-axis~\cite{Waser2015} and a double-$q$ structure with a superposition of (0, $\pi$) and ($\pi$,0) that maintains the tetragonal symmetry~\cite{Allred2016,Fernandes2015,Wang2015,Lorenzana2008,Gastiasoro2015,Avci2014,Mallett2015EPL}. This tetragonal AF (t-AF) order competes more strongly with SC than the o-AF one and leads to a $T_c$ reduction and a strong suppression of the SC condensate~\cite{Bohmer2015,Mallett2015PRL,Mallett2017PRB}.

The strong electronic correlations and, in particular, magnetic fluctuations have also pronounced effects on the band structure in the vicinity of the Fermi-surface (FS). In BKFA the FS is composed of several hole-pockets at the center of the Brillouin zone (BZ) ($\Gamma$-point) and electron pockets at the zone boundary (M-point) that have mainly Fe $xz$, $yz$, and $xy$ character. These bands are about two-times narrower than those predicted by DFT calculations~\cite{Qazilbash2009,Lu2008} and, in addition, are pushed toward the FS~\cite{Charnukha2015}. The latter effect arises from the strong particle-hole asymmetry and a pronounced interband scattering between the M and $\Gamma$-points that most likely involves AF fluctuations~\cite{Ortenzi2009,Cappelluti2011,Fanfarillo2016}. Especially near the M-point this yields very flat bands in the vicinity of the FS with a nearly singular behaviour that is very pronounced in the ARPES spectra of optimal doped Sm-1111 and BKFA~\cite{Charnukha2015,Evtushinsky2014}.

Although the bare electron-phonon coupling is commonly believed to be very weak~\cite{Boeri2008}, it can be strongly increased by the spin-phonon interaction~\cite{Boeri2010,Egami2010,Coh2016}.
This could possibly explain why a strong enhancement of the Fano effect of the in-plane $E_u$ Fe-As mode was observed in the o-AF state of undoped BaFe$_2$As$_2$~\cite{Akrap2009,Schafgans2011,Xu2018}. On the other hand, a sizeable Fano effect was also reported for near optimally doped BKFA without AF order~\cite{Xu2015,Yang2017}. To the best of our knowledge, a systematic study of the evolution of this Fano effect as a function of doping and temperature, and an assignment of the underlying electronic and magnetic excitations responsible for it, are still lacking.

\begin{figure*}[t]
\includegraphics[width=2\columnwidth]{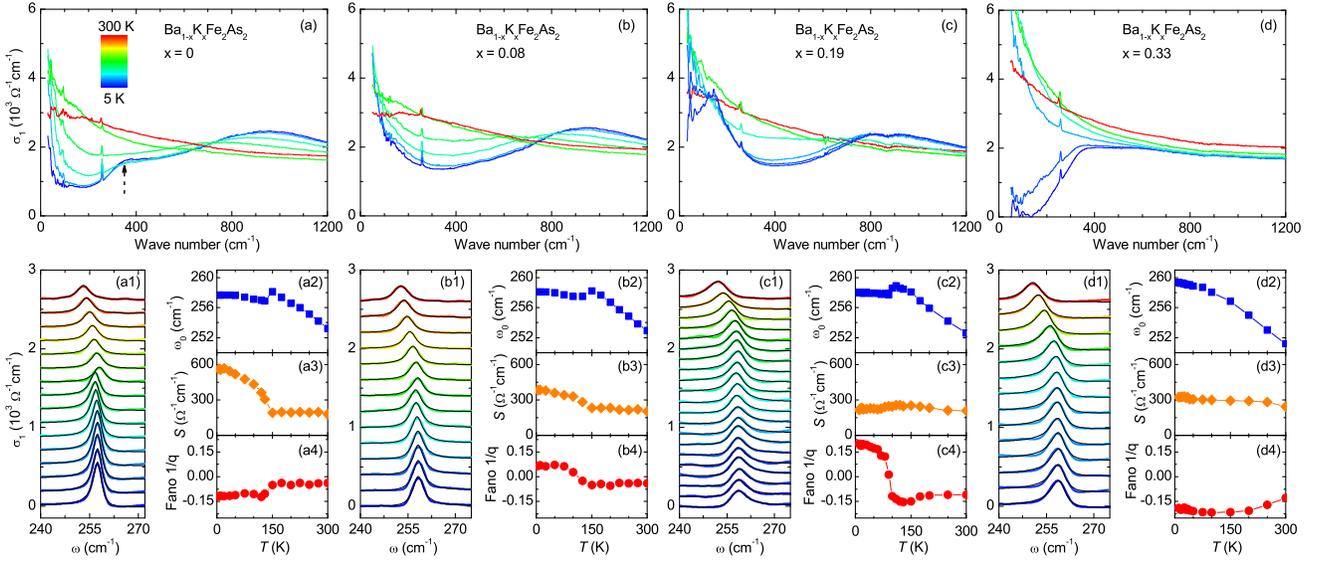}
\caption{ (color online) Temperature dependent optical conductivity of \BKFAx\ in the far infrared region for $x = 0$ (a), 0.08 (b), 0.19 (c) and 0.33 (d). (a1--d1) Line shape of the infrared-active phonon mode (with offset) at temperatures from 300 to 5~K (color lines). The underlying black solid lines through the data denote the corresponding Fano fit. Temperature dependence of the (a2--d2) phonon frequency $\omega_{0}$, (a3--d3) strength $S$ and (a3--d3) Fano parameter $1/q$ of the phonon for $x = 0$, 0.08, 0.19 and 0.33 in \BKFAx.}
\label{Fig1}
\end{figure*}
Here we close this knowledge gap by showing the detailed temperature and doping dependence of the Fano effect of the $E_u$ mode in BKFA. At low temperatures the strength of the Fano coupling, expressed in terms of the asymmetry parameter $1/q^2$, appears to be very sensitive to the magnetic and structural transitions at $x < 0.3$. Most remarkably, at temperatures well above these magnetic and structural transitions (in the paramagnetic tetragonal state), we find that $1/q^2$ exhibits a linear scaling with the SC critical temperature $T_c$. This striking observation is suggestive of an intimate relationship between the electronic excitations coupled to the $E_u$ phonon mode and the SC pairing mechanism.

%
%
A series of BKFA single crystals with $0 \leq x \leq 0.6$ were grown with a flux method~\cite{Karkin2014}. Their K-content was determined with X-ray diffraction and electron dispersive X-ray spectroscopy to an accuracy of ${\Delta}x \approx 0.02$~\cite{Mallett2015PRL,Mallett2017PRB}. The AF and SC transition temperatures, $T_N$ and $T_c$ were derived from the anomalies in the dc resistivity curves. The \emph{ab}-plane reflectivity $R(\omega)$ was measured at near-normal incidence with a Bruker VERTEX 70V spectrometer. An \emph{in situ} gold overfilling technique~\cite{Homes1993} was used to obtain the absolute reflectivity of the samples. The room temperature spectrum in the near-infrared to ultraviolet (4\,000--50\,000\icm) was measured with a commercial ellipsometer (Woollam VASE). The optical conductivity was obtained from a Kramers-Kronig analysis of $R(\omega)$~\cite{Dressel2002}. For the extrapolation at low frequency, we used the function $R = 1 - A\sqrt{\omega}$ (Hagen-Rubens) in the normal state and $R = 1 - A\omega^4$ in the SC state. On the high-frequency side, we used the room temperature ellipsometry data and extended them by assuming a constant reflectivity up to 12.5~eV that is followed by a free-electron ($\omega^{-4}$) response.

%
The upper panels of Figure~\ref{Fig1} show the temperature dependent spectra of the real part of the optical conductivity, $\sigma_1(\omega)$, in the infrared range for selected doping levels of $x$ = 0, 0.08, 0.19 and 0.33. Their infrared conductivity is dominated by the strong electronic response that is composed of a Drude peak at the origin, due to the itinerant carriers, and a pronounced tail toward high frequency, that arises from inelastic scattering of the free carriers and/or low-lying interband transitions~\cite{Benfatto2011,Charnukha2014,Marsik2013,Calderon2014}. The spectra agree well with the previously reported ones~\cite{Hu2008,Wu2010,Charnukha2013,Dai2013PRL,Xu2017,Mallett2017PRB} and show the well-known changes due to the spin-density-wave (SDW) at $T_N$ = 138~K, 130~K and 90~K for $x$ = 0, 0.08 and 0.19, respectively, and the SC gap below $T_c$ = 18~K, and 38~K at $x$ = 0.19 and 0.33, respectively. The SDW and the SC gaps both reduce the spectral weight of the regular charge carrier response. For the former this spectral weight is shifted to higher energy, where it forms a so-called pair-breaking peak, whereas in the SC state it is transferred to a $\delta(\omega)$ function at the origin that accounts for the infinite dc conductivity.

\begin{figure*}[tb]
\includegraphics[width=2\columnwidth]{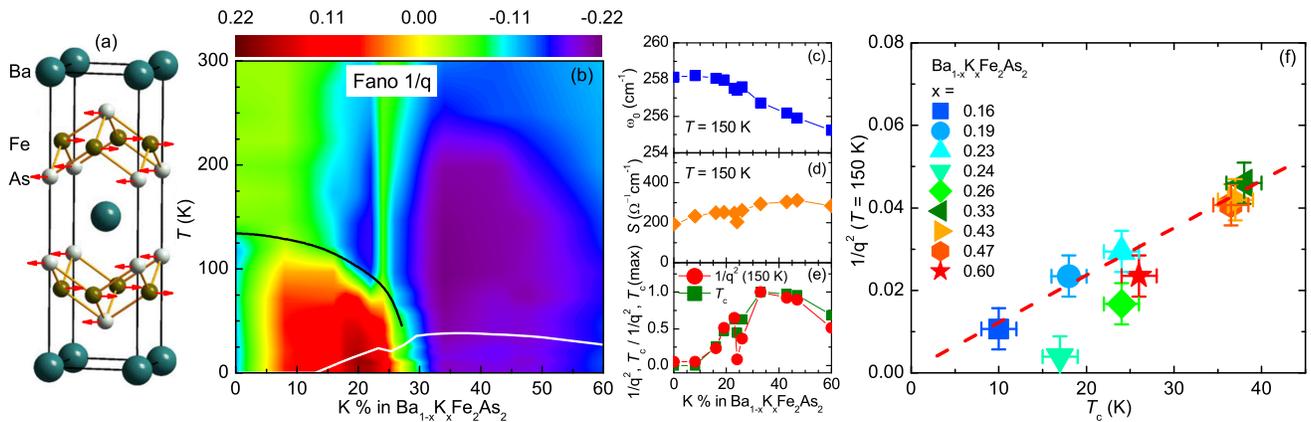}
\caption{ (color online) (a) Atomic displacements for the Fe-As stretching mode in BaFe$_2$As$_2$. (b) Color mapping of the Fano parameter $1/q$ in the $T$ versus $x$ plane. The solid black and white line denote the AF and SC transition temperatures, respectively. (c)--(d) Doping dependence of the phonon frequency $\omega_0$ and of the infrared strength $S$ of the $E_u$ phonon in \BKFAx\ in the normal state at 150~K. (e) Doping dependence of the Fano parameter $1/q^2$ at $T = 150$~K and of the SC transition temperature $T_c$ in \BKFAx. (f) Scaling of the Fano parameter $1/q^2$ vs. the SC transition temperature $T_c$ in \BKFAx.}
\label{Fig2}
\end{figure*}
In addition to these broad electronic features, the infrared-active Fe-As stretching mode (with $E_u$ symmetry)~\cite{Akrap2009} is clearly visible in all $\sigma_1(\omega)$ spectra as a weak but sharp mode around 255\icm. Its remarkable temperature and doping dependence is detailed on the left hand side of the lower panels of Fig.~\ref{Fig1} which show a magnified view of the Fe-As phonon mode with the electronic background subtracted. A sketch of the eigenvectors of the $E_u$ Fe-As phonon is displayed in Fig.~\ref{Fig2}a. For a quantitative analysis we fitted (solid lines in Figs.~\ref{Fig1}(a1-d1)) the optical properties of the $E_u$ mode with a Fano function~\cite{Fano1961,Cappelluti2010,Cappelluti2012}:
\begin{equation}
\label{Fano}
\sigma_{1}(\omega) = S\left[\frac{q^2 + 2qz -1}{q^2 (1 + z^2)}\right],
\end{equation}
where $z = (\omega - \omega_0)/\Gamma$, and where $\omega_0$, $\Gamma$ and $S$ are the frequency, linewidth and strength of the phonon mode, respectively. An asymmetric profile of the phonon lineshape is ruled here by the Fano parameter $1/q^2$, which reflects the strength of the coupling between the phonon and the underlying electronic excitations. When the electronic excitations are lacking or they are not coupled with the phonon resonance we have $1/q^2 = 0$ and a symmetric Lorentzian lineshape is recovered.

The temperature dependence of the so-obtained phonon parameters $\omega_0$, $S$ and $1/q$, is shown in the lower right panels of Fig.~\ref{Fig1}. For all magnetic samples, the combined AF and structural transition into the o-AF state gives rise to clear anomalies in the $T$-dependence of $\omega_0$, $S$ and, especially, of $1/q$. In the following we focus on the evolution of the latter. At $x$ = 0, in agreement with previous reports~\cite{Xu2018,Chen2017,Schafgans2011}, $1/q$ has a very small negative value in the paramagnetic state that increases strongly in magnitude below $T_N$ (Fig.~\ref{Fig1}a4). For the doped samples with $x$ = 0.08 and 0.19 the o-AF transition gives rise to corresponding anomalies, except that $1/q$ increase from a negative value above $T_N$ (that is larger at $x$ = 0.19 than at 0.08) to a large {\em positive} value below $T_N$. The origin of the abrupt sign reversal of $1/q$ is further discussed in the Supplemental Material~\footnotemark[1]. Finally, for the optimally doped sample without any AF order ($x$ = 0.33, Fig.~\ref{Fig1}d4) $1/q$ has the largest negative value and it is only weakly temperature dependent. Quite remarkably, there is hardly a signature of the SC transition at $T_c$ in the temperature dependence of the phonon parameters. This is a clear indication that the Fano effect of the $E_u$ Fe-As mode does not arise from the coupling with the itinerant charge carriers, for which the spectral weight in the vicinity of the phonon mode decreases below $T_c$ due to the formation of the SC energy gap~\cite{Li2008}. This implies that the Fano effect of this $E_u$ phonon mode is governed by the coupling to some interband transitions that are part of the electronic background at higher frequency (that is only weakly affected by the SC transition as shown in Fig.~\ref{Fig1}d).

The full doping and temperature dependence of $1/q$ is summarized as a color map in Fig.~\ref{Fig2}b. Also shown is the evolution of $T_N$ (solid black line), which is accompanied by abrupt changes of $1/q$, and of $T_c$ (solid white line) which hardly affects the $1/q$ value, as was already discussed above. In the following we focus on the doping dependence of $1/q$ without the influence of the AF and structural transitions as derived from the constant-temperature cut in the paramagnetic/tetragonal phase at $T$ = 150~K. Whereas the phonon frequency and intensity in Fig.~\ref{Fig2}c,d show only a weak doping dependence, the Fano parameter in Fig.~\ref{Fig2}e reveals a characteristic dome-like profile that closely resembles the one of the $T_c$ value and reproduces even the indentation at $x$ = 0.24 -- 0.26 due to the strong competition of SC with the t-AF order. This striking dome-like doping dependence of the Fano parameter persists up to room temperature (as is further detailed in the Supplemental Material)~\footnotemark[1].

The almost linear correlation in the doping range $x$ = 0 -- 0.6 between the Fano parameter of the $E_u$ mode in the paramagnetic state and the SC transition temperature is further highlighted in Fig.~\ref{Fig2}f where we plot $1/q^2$ versus $T_c$. Clear deviations from a linear behavior occur only for the samples with $x$ = 0.24 and 0.26, for which the competition with the t-AF phase causes an anomalous suppression of superconductivity and where additional complex physics is probably at work (e.g. due to a residual o-AF phase the $T_c$ value might be overestimated).

To shed more light on the physical processes responsible for the Fano effect of the $E_u$ Fe-As mode in these materials, we analyzed the optical data within the charged-phonon scheme that was originally developed for carbon-based materials,~\cite{Rice1992,Kuzmenko2009,Cappelluti2010,Cappelluti2012}. In this context the Fano asymmetry parameter, $1/q$, is essentially ruled by the imaginary part of a complex function $\chi(\omega)$ that can be identified as a dynamical response function between the current operator and the electron-phonon operator relative to the $E_u$ phonon. In particular, $1/q$ scales as the imaginary part at the phonon frequency $\omega_0$, i.e. $1/q \propto \mbox{Im}\chi(\omega_0)$, which is different from zero whenever the $E_u$ phonon is coupled to an electronic interband transition at $\omega_0$. To identify the relevant processes responsible for the Fano effect, we analyze the paramagnetic/tetragonal state without any structural distortion. Using the Slater-Koster approach~\cite{Daghofer2008,Calderon2009}, we consider a minimal tight-binding model containing the three orbitals ($xz$, $yz$, $xy$) relevant for the description of the low-energy band structure in the doping range under consideration. We compute the electron-phonon operator for the $E_u$ mode at linear order in the lattice displacement, and we analyze the properties of the corresponding charged-phonon response function $\chi(\omega)$. As detailed in the Supplemental Material~\footnotemark[1] we find that,  when only the $xz/yz$ orbital subsector is considered, the charged-phonon response function $\chi(\omega)$ is vanishing at the leading order, implying a corresponding vanishing Fano effect. This property is a consequence of the  underlying symmetries of the $xz/yz$ subsystem. On the other hand, a finite Fano asymmetry is possible when the $xy$ orbital component is taken into account, allowing for a finite contribution to $\chi(\omega)$  of interband particle-hole transitions between the $xy$ and $xz/yz$ components of the bands. Even though a quantitative estimate of $\chi(\omega_0)$ at the $E_u$ phonon frequency is beyond the scope of the present manuscript, we can nonetheless conclude that the contribution of the $xy$ orbital to the low-energy optical transitions plays a dominant role in the experimentally observed, dome-like doping dependence of the Fano asymmetry.

These conclusions hold true as long as the above-mentioned symmetries are preserved. In this respect, any breaking of the $xz/yz$ equivalence due e.g. to electronic or structural nematic phases can prompt additional low-energy transitions leading to an otherwise forbidden Fano-type coupling with the $E_u$ mode. We believe that this is the case in particular for the AF ordered states. Note, on the other hand, that the SC phase does {\em not} modify the symmetries ruling the charged-phonon response, explaining the observed lack of a significant change of the $E_u$ mode in the SC state.

The primary role of the $xy$ orbital character in understanding the phase diagram of the Fano effect of the $E_u$ mode, and the remarkable correlation between the Fano parameter $1/q^2$ and the SC critical temperature $T_c$ reported in  Fig.~\ref{Fig2}f, shed new light on the current understanding of the SC phase diagram as well. Whereas it seems unlikely that the electron-phonon coupling can be the primary mechanism of the SC pairing, the observed correlation between $1/q^2$  and $T_c$ suggests that both the Fano effect of the $E_u$ mode and the SC pairing interaction require the presence of a sizeable $xy$ orbital component.

This calls for a closer look into the band structure and its doping evolution. As discussed above, in the paramagnetic/tetragonal state the Fano effect of the $E_u$ mode arises primarily from low-energy transitions between the $d_{xy}$ and the $d_{xz/yz}$ bands. Such optical transitions at low energy occur only close to the M points, where recent ARPES data~\cite{Charnukha2015,Evtushinsky2014} showed that as doping increases the Fermi level is pushed  towards the bottom of the electron-like bands, so that the Fermi surface at the M points evolves from an elliptical shape to electron-like propellers. In this regime the bands close to  M points are remarkably flat, leading to a substantial increase of the density of states. This can in turn affect both the strength of the AF spin-fluctuation exchange between the M and $\Gamma$ pockets, that is responsible for the pairing glue, and the joint density of states of the optical interband transitions near the M point, that govern the Fano effect. The flattening of the $xy$ bands at the M point can also explain a sizeable spin-fluctuation exchange between the M pockets~\cite{Kreisel2017}. This can then provide a secondary $d$-wave pairing channel with respect to the dominant $s_\pm$ one, that manifests below $T_c$ with the appearance of excitonic-like Bardasis-Schrieffer resonances in the Raman spectra~\cite{Maiti2016,Bohm2018}. Moreover, it remains to be understood whether the $E_u$ phonon is just an ``accidental witness'' of the SC pairing interaction (via its sensitivity to the $xy$ orbital content at the Fermi-level) or whether it even plays a cooperative role and enhances $T_c$.

%
%
In summary, we performed a systematic study of the Fano effect of the in-plane FeAs stretching mode in Ba$_{1-x}$K$_{x}$Fe$_{2}$As$_{2}$. Firstly, we showed that the Fano effect of this phonon mode is strongly enhanced by the magnetic/structural transition into the orthorhombic AF state. Secondly, and most importantly, we observed a striking, linear relationship between the Fano parameter $1/q^2$, as measured at temperatures well above the magnetic/structural transitions, and the SC critical temperature, $T_c$.  Theoretical calculations based on symmetry considerations show that the $xy$ orbital component of the low-energy bands near the M-point of the BZ plays a central role for the Fano effect of the $E_u$ phonon mode. This calls for a detailed investigation of the role played by the same orbital degrees of freedom on the orbital-selective pairing mechanism based on spin-fluctuations exchange, and their possibly cooperative interplay with electron-phonon coupling.

%
%
Work at the University of Fribourg was supported by the Schweizer Nationalfonds (SNF) by Grant No. 200020-172611. L.B. acknowledges financial support by the Italian MAECI under the Italian-India collaborative project SUPERTOP-PGR04879. Work at IOP was supported by MOST of China (Grant No. 2017YFA0302903) and NSFC of China (Grant No. 11774400).

%
%

\end{document}